%% file: main.tex
\documentclass[sigconf]{acmart}

\usepackage{graphics}
\usepackage{graphicx}
\graphicspath{ {figures/} }

\usepackage{booktabs}
\usepackage{xspace}
\usepackage{amsmath}
\usepackage{bm}
\usepackage{multirow}
\usepackage{xcolor}
\PassOptionsToPackage{hyphens}{url}

\usepackage{url}

\copyrightyear{2021}
\acmYear{2021}
\setcopyright{acmlicensed}\acmConference[RAID '21]{24th International Symposium on Research in Attacks, Intrusions and Defenses}{October 6--8, 2021}{San Sebastian, Spain}
\acmBooktitle{24th International Symposium on Research in Attacks, Intrusions and Defenses (RAID '21), October 6--8, 2021, San Sebastian, Spain}
\acmPrice{15.00}
\acmDOI{10.1145/3471621.3471844}
\acmISBN{978-1-4503-9058-3/21/10}

\begin{document}

\settopmatter{printacmref=true}
\fancyhead{}


\date{}

\newcommand{\etal}{\textit{et al}.}
\newcommand{\eg}{e.g.,~}
\newcommand{\ie}{i.e.,~}
\newcommand{\etc}{etc}

\title{Marked for Disruption: Tracing the Evolution of Malware Delivery Operations Targeted for Takedown}


\author{Colin C. Ife$^{\star}$, Yun Shen$^{\dagger}$, Steven J. Murdoch$^{\star}$, and Gianluca Stringhini$^{\ddagger}$}
\affiliation{
  \institution{$^{\star}$University College London, $^{\dagger}$Norton Research Group, $^{\ddagger}$Boston University}
  \country{$^{\star\dagger}$United Kingdom, $^{\ddagger}$United States}
}
\email{{colin.ife,s.murdoch}@ucl.ac.uk, yun.shen@nortonlifelock.com, gian@bu.edu}

\begin{abstract}
\input{abstract}
\end{abstract}



\citestyle{acmnumeric}

\maketitle


\input{introduction}
\input{related}
\input{dataset}

\input{background-operations}
\input{methodology}

\input{analysis}
\input{discussion}

\input{conclusion}

\begin{acks}
\input{acknowledgements}
\end{acks}

\bibliographystyle{ACM-Reference-Format}
\bibliography{biblio}



\end{document}

%% file: abstract.tex
The malware and botnet phenomenon is among the most significant threats to cybersecurity today.
Consequently, law enforcement agencies, security companies, and researchers are constantly seeking to disrupt these malicious operations through so-called \textit{takedown} counter-operations.
Unfortunately, the success of these takedowns is mixed.
Furthermore, very little is understood as to how botnets and malware delivery operations respond to takedown attempts.
We present a comprehensive study of three malware delivery operations that were targeted for takedown in 2015--16 using global download metadata provided by Symantec.
In summary, we found that: 
(1) Distributed delivery architectures were commonly used, indicating the need for better security hygiene and coordination by the (ab)used service providers. 
(2) A minority of malware binaries were responsible for the majority of download activity, suggesting that detecting these ``super binaries'' would yield the most benefit to the security community. 
(3) The malware operations exhibited \textit{displacing} and \textit{defiant} behaviours following their respective takedown attempts. We argue that these ``predictable'' behaviours could be factored into future takedown strategies. 
(4) The malware operations also exhibited previously undocumented behaviours, such as \texttt{Dridex} dropping competing brands of malware, or \texttt{Dorkbot} and \texttt{Upatre} heavily relying on upstream dropper malware. These ``unpredictable'' behaviours indicate the need for researchers to use better threat-monitoring techniques.

%% file: introduction.tex

\section{Introduction}
\label{sec:disrupting:intro}



Malware delivery has evolved into a major business for the cybercriminal economy and a complex problem for the security community.
The \textit{botnet} -- a network of malware-infected devices that is controlled by a single actor through one or more command and control (C\&C) servers -- is one phenomenon that has benefited from the malware delivery revolution.
Diverse distribution vectors have enabled such malicious networks to expand more quickly and efficiently than ever before.
Once established, these botnets can be leveraged to commit a wide array of secondary computer crimes, such as data theft, financial fraud, coercion (ransomware), sending spam messages, distributed denial of service (DDoS) attacks, and unauthorised cryptocurrency mining~\cite{stringhini2014harvester,stone2011underground,kaspersky_botnet_2018,binsalleeh2010analysis,abu_rajab_multifaceted_2006}.
Even worse, these botnets could be further monetised as \textit{pay-per-install} services~\cite{caballero2011measuring}, allowing the operator to rent out access of their network to other criminals.
Finally, botnet operators employ a myriad of techniques to avoid detection and improve the resiliency of their operations, such as using software polymorphism~\cite{bayer2009view} to beat antivirus engines, Fast Flux Service Networks (FFSNs)~\cite{holz2008measuring} to rapidly change the IP addresses of their servers, Domain Generation Algorithms (DGAs)~\cite{antonakakis2012throw} to constantly change the domain names of their C\&C servers, and distributed servers spanning multiple regions for redundancy and elusive software delivery~\cite{Rossow:2013,lever2017lustrum}.


Because of the serious and growing threat that botnet and malware delivery operations pose to society, law enforcement agencies (LEAs), security companies, and researchers constantly seek methods, opportunities, and intervention points to disrupt such malicious operations~\cite{nadji2013beheading,edwards2015analyzing}.
Takedown operations are just a subset of some of the disruptive techniques employed: infiltrating botnets for intelligence-gathering and sabotage; re-routing network traffic meant for known C\&C servers to disrupt their communication channels (\ie a DNS sinkhole); forcing Internet service providers (ISPs) to shutdown malicious servers that they host; or physically seizing malicious server infrastructure and assets, and arresting the miscreants involved. 
The success of these operations is mixed~\cite{edwards2015analyzing}.

Although the details of a number of takedown operations have been recorded in the literature, few studies examine how the targeted malware delivery operations actually respond to such interventions.
This leaves many important questions unanswered:
After a takedown operation, what happens next?
Do the malware operations break down?
If not, how quickly do they resurface?
Do the operators move their infrastructure elsewhere, or perhaps change their modus operandi?
Assessing the efficacy of takedown operations, are there more effective intervention points in these malicious infrastructures?
Finally, considering the behaviours of these miscreants, could some of their reactions be predicted and taken into account by LEAs and security practitioners?

In this study, using global download metadata collected in 2015--2016, we devise a novel tracking and analysis methodology to quantitatively assess the global activity of malware delivery operations targeted for takedown, and how this activity evolves over the course of a year in light of these actions.
In particular, we focus on three malware delivery operations (botnets) that were targeted in the fall of 2015: the \texttt{Dridex}, \texttt{Dorkbot}, and \texttt{Dyre-Upatre} operations.
These botnets were selected as they were among the few known to have been targeted for takedown between October 2015--September 2016, corresponding to the collection period of the dataset used herein.
We analyse the activities of two operational components over the course of a year: the upstream server infrastructure (server-side), and the downloaded binaries and their dropper networks (client-side).
In summary, this study makes the following contributions:
\begin{enumerate}
    \item We devise a novel methodology to track and analyse malware delivery operations over time using download metadata.
    This methodology could be used to analyse any class of software delivery operation at scale, such as malware, potentially unwanted programs (PUPs), or benignware.
    \item We observe a myriad of behavioural responses to takedown attempts by each malware delivery operation.
    Specifically, we find that: (1) The use of distributed delivery architectures was common among the studied malware.
    (2) A minority of malware binaries were responsible for the majority of download activity.
    (3) The malware operations exhibited some ``predictable'' behaviours following their respective takedown attempts such as \textit{displacement}~\cite{hesseling_displacement_nodate} and \textit{defiance}~\cite{sherman_defiance_1993} behaviours.
    (4) The malware operations also exhibited previously undocumented behaviours, indicating the need for the research community to use better monitoring techniques.
\end{enumerate}
This study gives the security community deeper insight into the dynamics and complexities within malware delivery operations, particularly in light of a takedown attempt, while also uncovering challenges and further opportunities to disrupting such operations.

%% file: related.tex

\section{Related Work}
\label{sec:disrupting:related}

Botnet takedowns are counter-operations to disrupt botnet activities and the malware delivery networks that enable their growth.
There are diverse techniques to taking down botnets: botnet infiltrations~\cite{stone2009your}, ISP takedowns~\cite{carpou2015robots}, DNS sinkholes~\cite{stone2009your}, and arrest and seizure.
However, the fundamental problem with botnet takedowns is that if the botnet is not taken down fully or its operators not prosecuted, the operators may simply revive their operations and make them more resilient, making the task of taking down the botnet more difficult the next time round.
Because of this, various studies have attempted to quantify the effects of takedowns.
Clayton~\cite{clayton2009much} examined email statistics from a medium-sized UK ISP to assess the effects of the 2008 McColo takedown on global spam volume. It was found that significant reductions in spam email volumes around the time of the takedown operation.
However, it was also found that particular types of spam detection mechanisms employed by this ISP ceased to be as effective.
Dittrich~\cite{dittrich2012so} conducted a broader study, qualitatively analysing a set of highly publicised botnet takedown efforts between 2009-2011. It was concluded that, while some takedown strategies are more effective than others, the arms race between security practitioners and cybercriminals will continue to make botnet takedowns more expensive and difficult as cybercriminals will continue to make their infrastructures more resilient. The author called for more coordination and shared knowledge between the security community to make botnet takedowns more efficient and sustainable.

In an attempt to bring measurement and order to botnet takedown analysis, a takedown analysis and recommendation system, \textit{rza}, is proposed by Nadji~\etal~\cite{nadji2013beheading}.
This system allows researchers to conduct a post-mortem analysis of past botnet takedowns, and provide recommendations on how to execute future ones successfully.
This work is motivated with some real case studies.
In a second work, Nadji~\etal~\cite{nadji2015still} propose improvements to the \textit{rza} system by enhancing its risk formula to include botnet population counts.
Two additional botnet takedowns are also examined, and the policy ramifications of takedowns are discussed in detail by the authors.
Lerner~\cite{lerner2014microsoft} also discusses regulatory and policy solutions to botnet takedowns, particularly arguing the need for more public-private partnerships to achieve this endeavour.
Shirazi~\cite{shirazi2015botnet} surveys and taxonomises 19 botnet takedown initiatives between 2008--2014 and proposed a theoretical model to assess the likelihood of success for future botnet takedown initiatives. To the best of our knowledge, the author is still in the process of building this database before releasing it to the security community.

Investigating the effects of takedowns further, a recent historical study by Edwards~\etal~\cite{edwards2015analyzing} was conducted on the causal effects of botnet takedowns on ISPs that hosted spamming activity.
In this work, the authors build an autoregressive model for each ISP to model \textit{wickedness} -- a metric defined as total spam released per ISP -- as a function of (i) external factors and (ii) each takedown that occurred as represented as a time-lagged step-function.
They find that, for most takedowns, the effect of a takedown is minimal after a period of 6 weeks.
However, takedowns with a seizure element appear the most effective over the long-term.
They also find evidence of a takedown in one region causing a diffusion of benefits and crime in others.

%% file: dataset.tex

\section{Datasets}
\label{sec:dataset}

\subsection{Download Metadata}
\label{sec:data_source}
We use a download activity dataset provided by Symantec.
This anonymised dataset consists of download data from 12 million end-users of Symantec's products between October 1st, 2015 and September 29th, 2016. These users explicitly opted into the data-sharing programme, which does not include personally identifiable information.
These participating users periodically report metadata information on the binaries that they download, offering rich information regarding the time at which a binary is downloaded, which server it is downloaded from, and which program initiated the download activity. 
Equation~\ref{eq:download_event_2} outlines the structure of a download event:
\begin{equation}
    \label{eq:download_event_2}
    \mathbf{d} = <F_f,A_f,U_r,U_f,D,I,F_p,U_p>
\end{equation}
where $F_f$ is the downloaded file identified by its SHA-2; $A_f$ represents a set of attributes which provides additional information about file $F_f$, such as its filename, its size (in bytes), and its ``reputation'' and ``prevalence'' scores assigned by Symantec's analysis systems (see Section~\ref{sec:ground_truth});
$U_r$ is the initial (referrer) URL in an HTTP redirection chain; $U_f$ is the download URL (after removing URL parameters) while $I$ is the IP address of the download server and $D$ its fully qualified domain name (FQDN); $F_p$ is the SHA-2 of the parent file (or \textit{dropper}) and $U_p$ the source URL of the parent file.
In total, 81.5 million download events were observed over 53 days, which were sampled one day per week from October 1st, 2015.

Note that this study focuses on malicious file downloads.
To this end, we leverage the reputation scores assigned to files by Symantec, discard any file that has a high (benign) reputation score and is not confirmed as malicious by VirusTotal\footnote{VirusTotal is a free online service that analyses submitted files and URLs across different antivirus engines and website scanners.}, and only keep the files involved in the delivery of malware or PUP.
Note that we consider a file malicious if at least one of the top five AV vendors by market share (in no particular order, Avast, AVG, Avira, Microsoft, and Symantec) and a minimum of two other AVs detect it as malicious.
Other works have used a similar technique~\cite{Nelms2015webwitness,stringhini2017marmite,ife2019waves}.
We also filter out IP addresses that are not valid for public use as well aberrant data (e.g. events with no parent file or network resource).

\subsection{Binary Ground Truth}
\label{sec:ground_truth}
We utilise a variety of ground truth to establish whether files are malware or PUP, and to which families they belong.
This allows us to track the evolution of different malware and PUP delivery operations, especially in response to different mitigation strategies.

\noindent \textbf{Reputation and Prevalence Scores.}
Symantec employs extensive static and dynamic analysis systems to determine the maliciousness of a binary, as well as estimate its prevalence in the wild.


\noindent \textbf{VirusTotal.}
We query VirusTotal~\cite{virustotal} with each file SHA-2 to obtain the number of AV vendors that flag the file as malicious and the malware or PUP family labels designated to it by each vendor.
VirusTotal can sometimes take several months (or years) to flag malicious files in the wild due to coverage issues~\cite{lever2017lustrum,kwon2015dropper,peng2019opening}.
As such, this analysis is conducted 3--4 years after the data is first collected.
This makes sense since we seek maximise our ground-truth data (namely family labels) so as to characterise the evolution of different malware and PUP operations as accurately as possible.

\noindent \textbf{AVClass.}
In conjunction with VirusTotal, we utilise the AVClass malware labelling tool~\cite{sebastian2016avclass} to remove ``noisy'' and conflicting family labels for a given file so as to determine a correct and consistent one.
For example, multiple AV engines may generate labels of \texttt{Adware.Rotator.F}, \texttt{Adware.Generic}, and \texttt{Adware.Adrotator.\\Gen!Pac} for the same AdRotator SHA-2 (a PUP).
We utilised an updated set of AVClass family labels at the time of this study.\footnote{Commit 21806f3 from \url{https://github.com/malicialab/avclass} (July 27th, 2018)}

\noindent \textbf{National Software Reference Library.}
NSRL Reference Data Set (RDS) version 2.67 provides us SHA-2 hashes of known benign and reputable programs involved in malicious file delivery. 


%% file: background-operations.tex

\section{Targeted Malware Operations}
\label{sec:disrupting:background:targeted}
In this study, we seek to understand how malware operations evolve in light of takedown operations against them.
However, it is important to first identify takedowns that occurred within the dataset collection period, \ie~between 1st October, 2015 and 29th September, 2016.
We identify three different botnets that were targeted for takedown within the subject period: Dridex, Dorkbot, and the Dyre-Upatre malware delivery operations.

\subsection*{Dridex}
The Dridex malware (also known as Bugat, Cridex, Drixed, and Dridexdownloader) is a banking trojan and botnet malware, specifically designed to steal banking credentials and other personal information on a compromised system.
Dridex has been known to spread through phishing emails as a malicious attachment, to self-replicate by copying itself from compromised devices to mapped network drives and local storage devices~\cite{symantec_dridex_nodate}, and to be delivered through exploit kits on compromised web servers~\cite{malwarebytes_trojandridex_nodate}.
Between August and October, 2015, one of the botnet operators was arrested, while the NCA in the UK undertook a DNS sinkhole operation against Dridex servers.
Between 9th October and 8th December, 2015, LEAs conducted a second \textit{DNS sinkhole} and \textit{disinfection} campaign against Dridex, though the specifics are unknown~\cite{fbi_bugat_nodate,doj_bugat_2015}.

\subsection*{Dorkbot}
The Dorkbot malware is a family of worms known to steal data from compromised systems, disable security applications, and form botnets to distribute other types of malware~\cite{microsoft_win32dorkbot_nodate,certpl_dorkbot_2015}.
It has been known to propagate through infected USB flash drives, instant message applications, social networks, spam messages, and exploit kits.
Researchers identified Dorkbot diversifying its C\&C servers to multiple regions, such as throughout Europe, Asia, and North America~\cite{certpl_dorkbot_2015,theregister_dorkbot_nodate}.
Around 3rd December 2015, security companies and LEAs around the world conducted a swift \textit{DNS sinkhole} and \textit{seizure} operation against the Dorkbot botnet~\cite{welivesecurity_dorkbot_2015,microsoft_dorkbot_2015,welivesecurity_dorkbot_2015,theregister_dorkbot_nodate}.

\subsection*{Dyre and Upatre}
The Dyre and Upatre operations provide an interesting case study, not least given the reported LEA operation against the Dyre botnet coincided with a sudden, global drop in malicious download activity, which was observed in an earlier study~\cite{ife2019waves}.
Dyre (also known as Dyreza, Dyzap, and Dyranges) is a sophisticated financial fraud trojan that targets Windows computers~\cite{zemanacom_what_nodate}.
However, most notably, security researchers have identified the Dyre-Upatre relationship as being key to its operation, where, after hosts are infected with Upatre malware, it proceeds to install Dyre malware onto these devices~\cite{zemanacom_what_nodate,symantec_dyre_nodate,intelligence_dyre_nodate}.
More specifically, Upatre is a dedicated dropper malware: once on a victim machine, its sole purpose is to deliver additional malware components onto it.
However, besides delivering Dyre samples, Upatre has been known to distribute other malware families such as GameOver Zeus, Kegotip, Locky, and Dridex~\cite{noauthor_upatre_2018}.

In this study, we focus only on the activities of the Upatre dropper as little to no observable Dyre download activity is found in this dataset.
Why exactly this is the case is not known.
Since Dyre was known to undergo rapid polymorphism~\cite{zemanacom_what_nodate}, it could be indicative of the inability of antivirus engines to keep up with its high churn of malware binaries, or some form of measurement error with the telemetry sensors used to collect this dataset.
Around 18th November 2015, law enforcement officials conducted a \textit{seizure} and \textit{arrest} operation against the Dyre operators in Moscow, Russia~\cite{brewster_russian_nodate}.

%% file: methodology.tex

\section{Methodology}
\label{sec:disrupting:methodology}

Although the principal focus of this study is the dynamics of three specific malware delivery operations targeted for takedown, we devise a methodology that may be adopted to analyse any class of file delivery operation, whether malicious or benign.
Therefore, in this section, we detail the steps to (i) build the download graphs for the year-long dataset; (ii) classify the file nodes as either malware, potentially unwanted programs (PUPs), or benign, along with their specific software brands/families; and (iii) aggregate and track each software delivery operation in time, with a particular focus on their evolving use of \textit{delivery infrastructure} and their \textit{dropping behaviours}.

It is pertinent to note that, in this study, we only seek to analyse file delivery \textit{operations} -- not file delivery \textit{campaigns}.
More precisely, we only analyse aggregate (global) file delivery activity pertaining to a given software family (\eg all \textit{Zeus} malware delivery activity).
This is opposed to the more fine-grained analysis of individual clusters of activity (campaigns) pertaining to a single software family (\eg individual \textit{Zeus} botnet campaigns, which may involve independent operators by virtue of its crimeware-as-a-service business model~\cite{sood_crimeware-as--servicesurvey_2013}).
We align with the above distinction between the terms \textit{operation} and \textit{campaign} for the purposes of this study.
As such, disentangling individual delivery campaigns (and the respective actors) for a given operation is beyond the scope of this study.

\begin{figure}[t]
    \centering
    \includegraphics[width=0.8\linewidth]{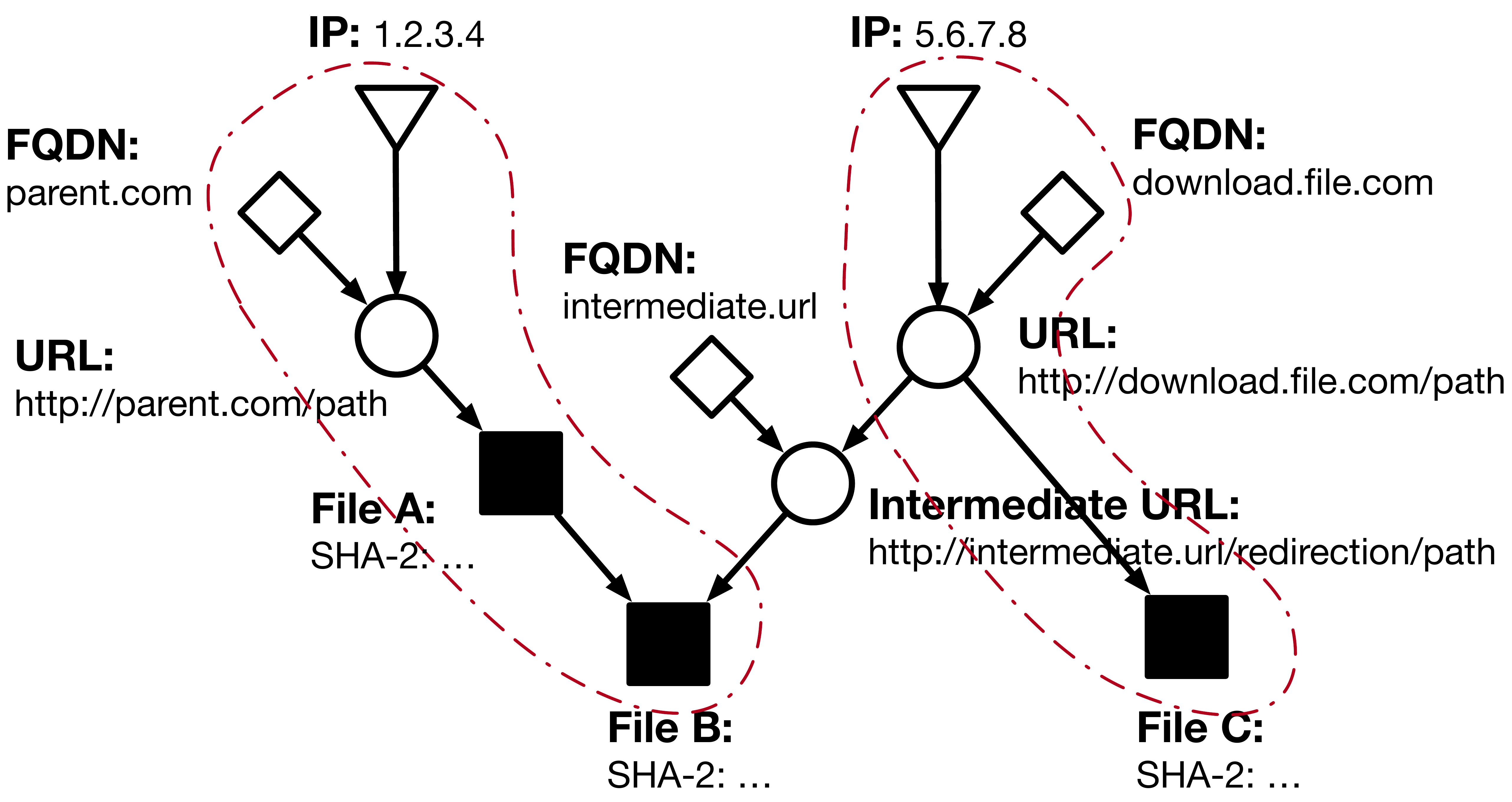}
    \caption{A legend to interpret download graphs, adapted from~\cite{ife2019waves}. Two series of download events are highlighted.}
    \label{fig:download_graph}
\end{figure}

\subsection{Building Download Graphs}
\label{sec:disrupting:methodology:download_graphs}
We adopt the graph-building methodology used in similar work~\cite{ife2019waves}.
In summary, we build a directed graph $G=(V,E)$ where $V$ is a vertex list representing different entities (file SHA-2s, URLs, IPs, and FQDNs) and $E$ is an edge list representing relationships between nodes from the same download events.
An example of this graph schema is shown in Figure~\ref{fig:download_graph}, which includes FQDNs.
Taking download event 1, for example, \texttt{File A} is a dropper that was downloaded from \texttt{parent.com/path} with IP \texttt{1.2.3.4}. \texttt{File A} initiates the download of \texttt{File B}, it first makes a request to \texttt{intermediate.url}, which redirects to the terminal URL \texttt{download.file.com/path}.

\subsection{File Classification}
\label{sec:disrupting:methodology:file_classification}
Having constructed the download graphs for each observation window, we build on the file classification technique used in previous work~\cite{ife2019waves}. Specifically, we label each file node (based on its SHA-2) as either \textit{malware}, \textit{PUP}, or \textit{benign} using the ground truth sources outlined in Section~\ref{sec:ground_truth}, or leave it as \textit{unlabelled}.
If known, we also specify the \textit{software family} to which the SHA-2 belongs, whether malicious or benign.
Otherwise, we label SHA-2s without known family labels as \textit{singletons}.
In total, we classify 1,034,763 malicious file SHA-2s (4.83\% of all files), 443,541 (2.07\%) of which are classified as malware, and the remainder as PUP.
On the other hand, 350,517 SHA-2s (1.64\%) are known to be benign, as either VirusTotal flags them as not malicious (349,746 files), and/or the NSRL maintains that they are reputable (9,007 files).


\subsubsection{Aggregating Family Aliases}
A major part of this study is to analyse the activities of three malware delivery operations: Dridex, Dorkbot, and Upatre.
It is common for some antivirus engines to label each malware family differently, which may lead to multiple aliases being observed that refer to the same malware family.
Therefore, we configure the AVClass tool to map specific aliases to specific families.
Specifically, based on the sources for each malware operation in Section~\ref{sec:disrupting:background:targeted}, we aggregate the following aliases:
\begin{itemize}
    \item Dridex, Cridex, Bugat, Drixed, Dridexdownloader $\xrightarrow{}$ \texttt{Dridex};
    \item Dorkbot, Ngrbot $\xrightarrow{}$ \texttt{Dorkbot}; and
    \item Upatre $\xrightarrow{}$ \texttt{Upatre}.
\end{itemize}
Other known aliases for these families that are ambiguously designated (\ie used to refer to several, independent malware families) or were not observed in the dataset were omitted.

\begin{table*}[t]
\centering
\resizebox{0.55\linewidth}{!}{
\begin{tabular}{ll}
\hline
\textbf{Metric} & \textbf{Description} \\
\hline
\hline
\multicolumn{2}{r}{\textit{Aggregate Network Activity$^{\star}$}} \\
\cline{1-2}
URL count & Total no. of URLs used in file delivery.\\
FQDN count & Total no. of FQDNs used in file delivery.\\
E2LD count used & Total no. of e2LDs used in file delivery.\\
IP count & Total no. of IP addresses used by file delivery servers. \\
Country count & Total no. of countries associated with file delivery servers.\\
\cline{1-2}
\multicolumn{2}{r}{\textit{Evasion Indicators$^{\star}$}} \\
\cline{1-2}
IP count per e2LD used & No. of IPs associated with each e2LD used in file delivery.\\
E2LD count per IP used & No. of e2LDs associated with each IP used in file delivery. \\
\hline
\multicolumn{2}{r}{\textit{Aggregate Download Activity$^{\dagger}$}} \\
\cline{1-2}
Download count & Total no. of times the target family is downloaded.\\
Drop count & Total no. of times the target family delivers other files.\\
Download count per SHA-2 & No. of times each target family SHA-2 is downloaded.\\
Drop count per SHA-2 & No. of times each target family SHA-2 delivers other files.\\
\cline{1-2}
\multicolumn{2}{r}{\textit{Relational Dynamics$^{\dagger}$}} \\
\cline{1-2}
Parent SHA-2 count & Total no. of SHA-2s used to deliver the target family.\\
Child SHA-2 count & Total no. of SHA-2s delivered by target family. \\
\cline{1-2}
\multicolumn{2}{r}{\textit{Distributed Delivery Indicators$^{\dagger}$}} \\
\cline{1-2}
URL count per SHA-2 & No. of URLs used to deliver each target family SHA-2.\\ 
IP count per SHA-2 & No. of IPs used to deliver each target family SHA-2. \\
E2LD count per SHA-2 & No. of e2LDs used to deliver each target family SHA-2.\\
\cline{1-2}
\multicolumn{2}{r}{\textit{Polymorphism Indicators$^{\dagger}$}} \\
\cline{1-2}
SHA-2 count & No. of target family SHA-2s observed.\\
SHA-2 churn & No. of SHA-2s in observation $i$ lost in observation $i+1$.\\
File size per SHA-2 & File size of each SHA-2 in kilobytes.\\
Reputation score per SHA-2 & Malice score assigned to each SHA-2 by Symantec.\\
Prevalence score per SHA-2 & Prevalence score assigned to each SHA-2 by Symantec. \\
& N.B: Prevalence indicates how often a SHA-2 is detected. \\
\hline


\end{tabular}
}
\caption{The network$^{\star}$ and downloader$^{\dagger}$ metrics used to analyse each malware delivery operation.}
\label{table:disrupting:metrics}
\end{table*}

\subsection{Tracking and Analysing Operational Activity}
\label{sec:disrupting:methodology:tracking_operation_activity}
Besides just monitoring malicious file presence, we want to establish how their use of delivery infrastructure and their dropping behaviours evolve alongside them.
It is particularly interesting to understand the evolution of malicious file delivery operations in the wake of different, disruptive strategies being applied against them, such as botnet takedowns or coordinated arrests.
We achieve this goal in two stages.
First, we devise a methodology to identify and track a (malicious) file delivery operation.
And second, we derive a set of metrics that describe different aspects of a given file delivery operation, and conduct time series analysis on these metrics.

\subsubsection{Tracking Delivery Operations}
Our approach to tracking delivery operations is simple:
For a (target) software family that we seek to analyse, $SF$, and for the $i$th observation period, where $i \in [1..53]$ (\ie every Thursday for a year), we conduct the following:
\begin{enumerate}
    \item We compute $F_{i}^{SF}$: the set of all file nodes pertaining to software family $SF$ in observation period $i$.
    \item We compute $P_{i}^{SF}$: the set of all parent nodes (URLs, IPs, FQDNs, parent files) involved in the download events that deliver the files $F_{i}^{SF}$ in observation period $i$.
    In terms of real-world actors, these parent nodes could be attributed to, for example, upstream hosting services, compromised websites, or pay-per-install network operators and affiliates~\cite{caballero2011measuring}.
    \item Likewise, we compute $C_{i}^{SF}$: the set of all child nodes (files) that are dropped by the files in $F_{i}^{SF}$ in observation period $i$.
    Being payloads, these child nodes could be attributed to the clients of the $SF$ delivery network. 
    \item Finally, we compute the node attribute look-up table, $A_{i}^{SF}$, which stores the attributes of all 
    nodes forming the delivery network of software family $SF$ in each observation period $i$ (\eg family, \# of downloads/drops, country, domain name).
\end{enumerate}

\subsubsection{Time Series Analysis}
\label{sec:disrupting:methodology:time_series}
We seek to generate a set of metrics (or features) which sufficiently describe the different aspects of a file delivery operation.
Using the data structures, $F^{SF}$, $P^{SF}$, $C^{SF}$, and $A^{SF}$ as defined above, we compute and analyse time series data based on two groups of metrics:

\noindent \textbf{Network dynamics}.
These metrics capture the dynamics of server-level activity in the file delivery operation.
The numbers of URLs, domains, IPs, and countries used to host delivery servers indicate the pervasiveness and extent of resources used for each operation.
The numbers of IPs associated with each domain provide indicators of use of the Fast Flux technique (rapidly changing IPs)~\cite{holz2008measuring}, or the use of content distribution networks (CDNs) and multi-region servers -- common methods used by botnet to avoid detection and increase resiliency~\cite{Rossow:2013}.
On the other hand, the number of domains associated to any given IP could be an indicator of use of shared-hosting clusters, or servers using domain generating algorithms (DGA) -- another commonly used technique by C\&C servers to avoid detection~\cite{plohmann2016comprehensive}.
Finally, we also quantify the most popular domains, 
IP addresses, and regions used in each operation. 

\noindent \textbf{Downloader dynamics}.
These metrics capture information relating to the software family in question and the binaries it uses to drive the delivery operation.
Specifically, we obtain the total and per-SHA-2 counts of download and dropping events for the software family -- key performance indicators of delivery operations.
We also keep track of the total and top $N$ families involved in the software family's download activities.
Further, we analyse the numbers of URLs, domains, and IPs used to deliver each file SHA-2, which are all indicative of the use of diverse distribution vectors, perhaps to increase outreach to end-users, or to evade detection systems more effectively.
We also examine metrics that indicate polymorphism (a malware characteristic to evade detection~\cite{bayer2009view}): the number of SHA-2s observed, their churn rates, and the distributions of their file sizes, malice (reputation) scores, and prevalence scores (as detected and assigned by Symantec).
It should be noted that a higher malice score corresponds to a higher likelihood that a file is malicious (see Section~\ref{sec:ground_truth}), while a higher prevalence score indicates that a file is detected more frequently.

These metrics are summarised in Table~\ref{table:disrupting:metrics}.
We analyse the time series derived from these metrics in Section~\ref{sec:disrupting:operation_analysis}.

\noindent \textbf{In summary,} this methodology and rich dataset grants us an unprecedented insight into the dynamics of malicious file delivery operations, the business relationships between them, and, most importantly, how they each react to disruptive counter-operations.

%% file: analysis.tex
\section{Analysis}
\label{sec:disrupting:operation_analysis}
In this section, we use the techniques described in Section~\ref{sec:disrupting:methodology:tracking_operation_activity} to analyse the evolution of three different malware delivery operations: the Dridex, Dorkbot, and Dyre-Upatre botnets.
Each botnet faced law enforcement takedown attempts between October 2015 and September 2016.
For each malware delivery operation, we mark the time period of the associated takedown operation and analyse the botnet's activities afterwards.
We focus our analysis on two types of metrics: the network dynamics, and the downloader dynamics.

\subsection{Network Dynamics}
\label{sec:disrupting:network_dynamics}
\begin{figure}[t]
    \centering
    \includegraphics[width=1.05\linewidth]{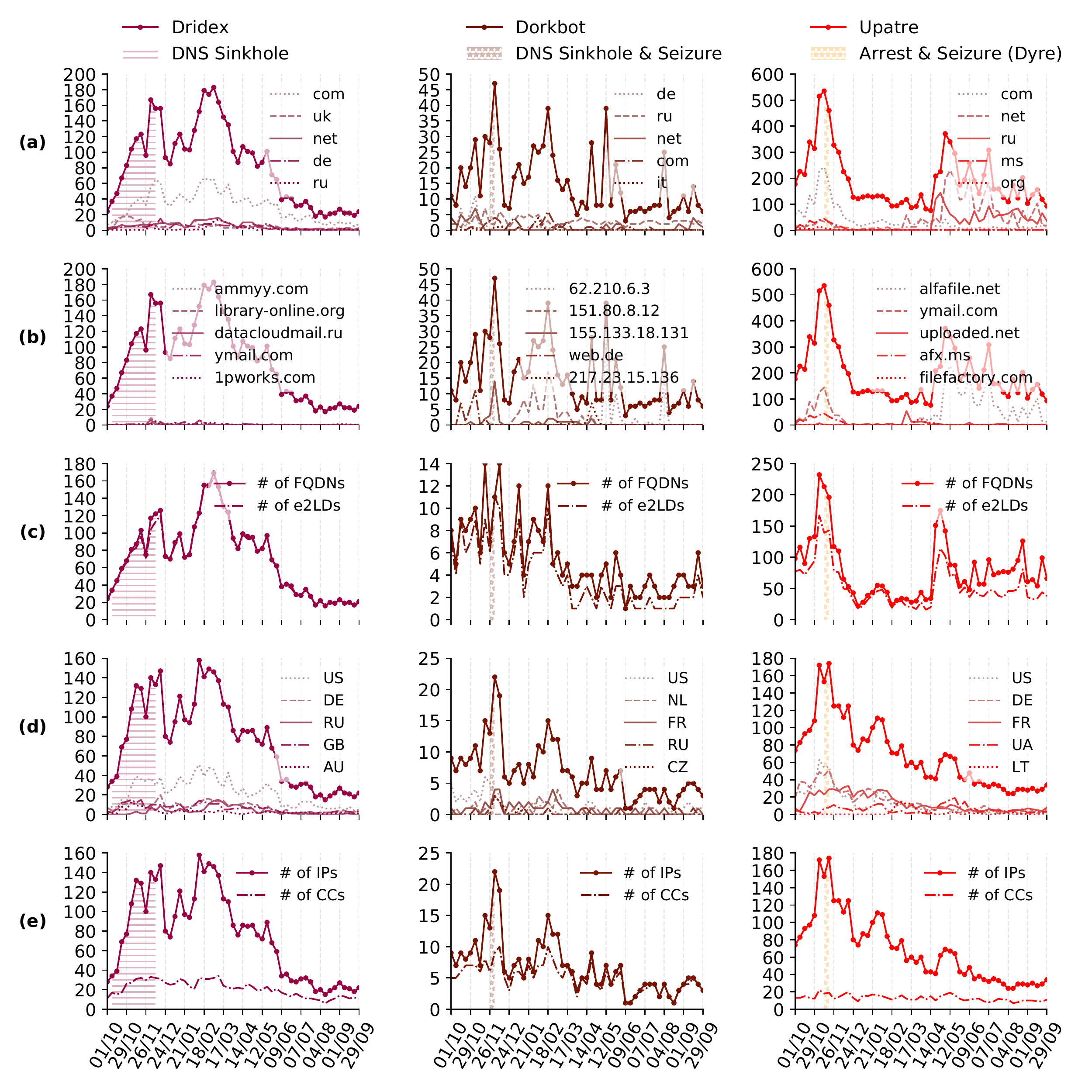}
    \caption{\textbf{Aggregate network activity:} \textbf{(a)} \# of URLs used and top 5 TLDs; \textbf{(b)} \# of URLs used and top 5 e2LDs/IPs; \textbf{(c)} \# of FQDNs and \# of e2LDs; \textbf{(d)} \# of IPs used and top 5 hosting countries; and \textbf{(e)} \# of IPs and \# of hosting countries.
    Dridex exhibits consistent growth in network activity during the DNS sinkhole, while Dorkbot and Upatre both exhibit significant, short-term drops in network activity after their respective takedowns with varying long-term responses.}
    \label{fig:disrupting:operations_network_dyn}
    \includegraphics[width=1.05\linewidth]{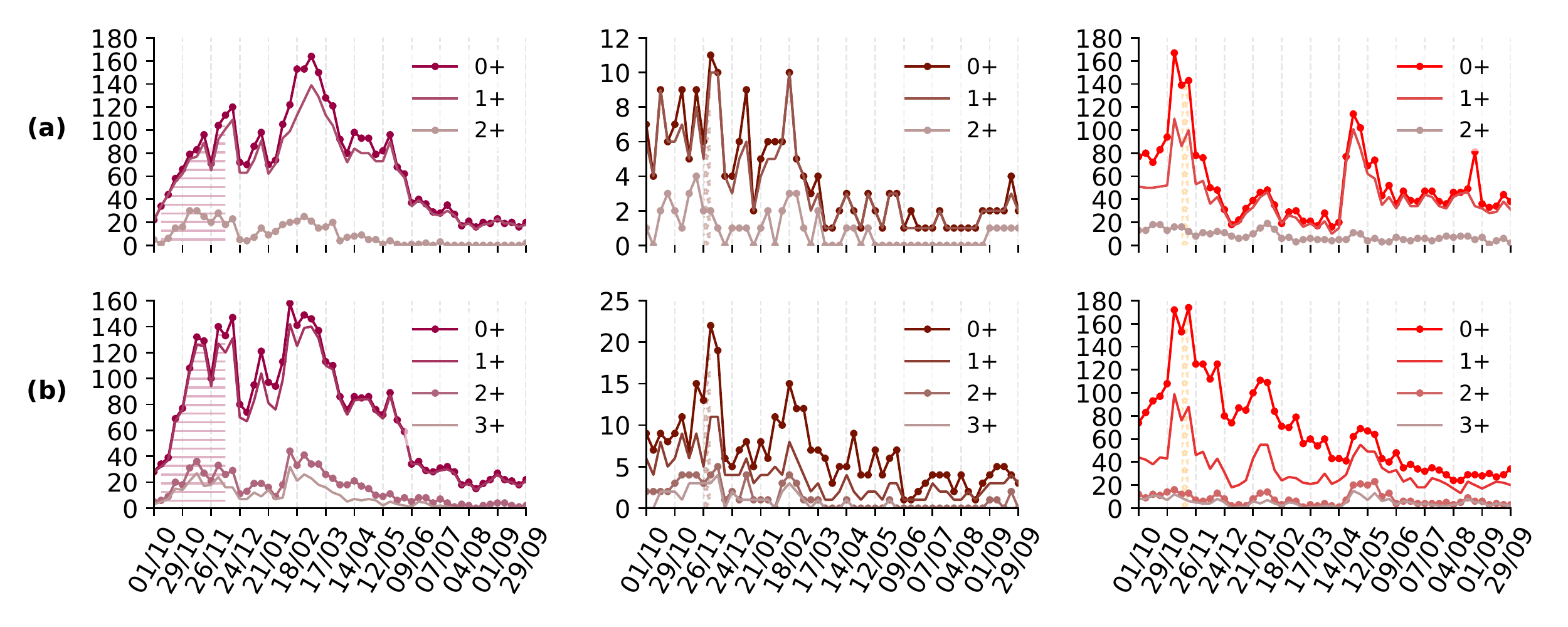}
    \caption{\textbf{Evasion indicators:} \textbf{(a)} \# of e2LDs associated with $N+$ IPs;  and \textbf{(b)} \# of IPs associated with $N+$ e2LDs. Dridex was found to use shared-hosting platforms and CDNs often. Upatre increases its use of IPs with $2+$ domains from mid-April, most of which were for \texttt{.ru} DGA domains.}
    \label{fig:disrupting:operations_network_dyn_2}
\end{figure}

We begin our analysis with the upstream infrastructure of each malware delivery operation. To this end, we compute and analyse the network dynamic metrics described in Section~\ref{sec:disrupting:methodology:time_series}.
Figure~\ref{fig:disrupting:operations_network_dyn} shows a number of time series denoting aggregate network dynamics, and Figure~\ref{fig:disrupting:operations_network_dyn_2} evasion indicators for each malware delivery operation.
We note some apparent features.
For instance, Dridex exhibits consistent growth in all forms of network activity from early October 2015 (despite the DNS sinkhole operation) until the end of February 2016, after which its network activities tail off.
On the other hand, the Dorkbot and Upatre operations (which faced ``seizure'' counter-operations) both exhibit significant drops in overall network activity in the short-term, with varying long-term responses.
This is consistent with the findings of other researchers~\cite{edwards2015analyzing} in that, though botnet responses to takedowns are highly variable, takedowns that involve the physical seizure of botnet infrastructure are usually associated with longer-lasting and more significant effects.

\subsubsection{Dridex}
Looking more deeply into the network dynamics of the Dridex operation, we see two distinct stages of activity.
Initially, there is a consistent increase and diversification of its server usage in all respects, specifically over the course of the 60-day DNS sinkhole counter-operation (1st October--3rd March).
This is followed by a stage of consistent decrease in network activity from 3rd March--29th September.
We find some interesting observations.

The first observation is that the period of consistent growth in malicious server activity seems to align with the 60-day period of the DNS sinkhole.
On the other hand, once this sinkhole operation concluded in early December 2015, Dridex server usage appears to fall and rise for a number of weeks.
Why this sequence of events occurs is unclear.
One would typically expect malicious server activity to decrease or to remain at a controlled level during a DNS sinkhole operation, as observed by other researchers~\cite{edwards2015analyzing}.
This is clearly not the case in this instance, where we observe the opposite.
Nonetheless, one must consider (at least) two factors regarding this observation.
First, the DNS sinkhole operation itself may not have been effected adequately or consistently.
It is possible that the Dridex operators switched to back-up or alternative servers that were not tracked by the agencies enforcing the sinkhole counter-operation.
At the same time, it is possible that the C\&C servers targeted for DNS sinkholing were separate to the servers used to deliver Dridex malware to the end-hosts.
If this was the case, this could highlight a significant limitation of DNS sinkholing as a sole countermeasure.
Second, it is likely that the Dridex operators were already aware of the impending law enforcement operation, taking into account the earlier arrest of one of their operators in August 2015, the preceding sinkhole operation by the National Crime Agency in September 2015, and the fact that the US authorities had already served four of the other Dridex operators notices of indictment~\cite{fbi_bugat_nodate}.
As a result, and perhaps in retaliation, the botnet operators may have increased their activities and/or moved their operations elsewhere, both of which would lead to an overall increase in network activity during this period.

The second observation of interest is that we see significantly increased usage of download URLs with a \texttt{.com} suffix 
(Figure~\ref{fig:disrupting:operations_network_dyn}(a)) and an increased usage of servers hosted in the US (Figure~\ref{fig:disrupting:operations_network_dyn}(d)).
Likewise, we see similar (albeit less significant) increases in URLs with \texttt{.uk} suffixes and GB-based servers.
Given that US law enforcement (along with that of the UK) were the driving force behind the Dridex takedown efforts, this increased usage of US-based (and to a lesser extent, GB-based) servers and domains could again be indicative of a concerted response by the Dridex operators.
Specifically, the malware operators could have been targeting US infrastructure and end-users in reaction to their takedown attempts.
At the same time, without any additional data, one cannot rule out the possibility that the Dridex operation had a significant dependence on US infrastructure prior to these takedown efforts, which would mean that they could just be attempting to recover lost ground.
Nonetheless, it is clear that these malware operators ramped up their operations at the same time that law enforcement were launching a counter-operation against them, culminating in significantly increased network activity over the ensuing months.

It is also interesting to note that the Dridex operation did not rely on any one download server or region.
This is shown by the low proportion of download activity via the most commonly used domains (\eg \texttt{ammyy.com}, \texttt{library-online.org} -- see Figure~\ref{fig:disrupting:operations_network_dyn}(b)).
This is also reflected in the approximate 1:1 ratio in \# of FQDNs-to-\# of e2LDs attributed to its download servers (Figure~\ref{fig:disrupting:operations_network_dyn}(c)).
Similarly, as Figure~\ref{fig:disrupting:operations_network_dyn}(e) shows, up to 35 different countries are used to host Dridex download servers.
Querying the data, we find that the Dridex operation makes significant use of (i) websites on shared-hosting platforms, and (ii) multi-region CDNs (such as \texttt{dropbox.com} or \texttt{googleusercontent.com}) as malware delivery vectors.
This accounts for the distributions of domains using multiple IPs and IPs using multiple domains (see Figures~\ref{fig:disrupting:operations_network_dyn_2}(a)--(b)).
This diversity in distribution channels makes it difficult to identify bottlenecks in the Dridex operation.
This could have been implemented by design, or a learned adaptation to previous takedown attempts.

Finally, as we see in the second era of its network activity, the Dridex operation appears to ``wind down'' its server usage just as quickly as it grew in the preceding months.
This reduced server usage seems to stabilise from around 4th August 2016.
Without additional data, it is difficult to draw any robust conclusion on what causes this reduction in activity (\eg whether it was a consequence of a takedown operation, or a conscious decision by the malware operators to reduce operational activity). 

\subsubsection{Dorkbot}
On a general note, the network activity of the Dorkbot operation appears to be varied and highly stochastic in clear contrast to the other malware operations.
It also appears that the Dorkbot operation is significantly less diverse in its use of download servers, as indicated by its use of fewer unique URLs, domains, and IPs in the Dorkbot delivery operation.
Further, we previously noted the sharp decline in Dorkbot's overall network activity just after the DNS sinkhole and seizure counter-operation.
However, due to its stochastic nature, it is difficult to determine the significance of this decline as Dorkbot exhibits erratic changes in network activity, both before and after the takedown operation.

Analysing its network dynamics more closely in Figures~\ref{fig:disrupting:operations_network_dyn}(a)--(b), Dorkbot's overall use of download/redirection URLs shows some cyclicity.
Specifically, we observe peaks in the number of URLs used roughly every 12 weeks.
A similar pattern is observed with its use of IPs, as shown in Figure~\ref{fig:disrupting:operations_network_dyn}(d)--(e), albeit with a more pronounced, downward trend.
It must be said that this pattern does not appear in Figure~\ref{fig:disrupting:operations_network_dyn}(c), which shows Dorkbot's (equally stochastic) use of domains gradually decaying for a few months before oscillating around a reduced baseline.
Looking at its use of top e2LDs/IPs in Figure~\ref{fig:disrupting:operations_network_dyn}(b), it is clear that these peaks in URL and IP activity are linked.
Particularly, the Dorkbot operation tends to rotate between specific server IPs to spearhead its network-based delivery activities: initially, it primarily uses \texttt{web.de} (a German TLD) between 1st October--12th November, then it briefly moves to \texttt{155.133.18.131} (a server in Poland) between 12th November--17th December, traversing the takedown period.
Afterwards, it moves to \texttt{151.80.8.12} (a server in France) from 24th December--31st March, before briefly switching to \texttt{217.23.15.136} (a server in Netherlands) from 31st March--5th May, before fluctuating in its use of \texttt{62.210.6.3} (a server in France) from 5th May--11th August.

This pattern of displacement in Dorkbot's server usage appears to be highly coordinated, although the cause or purpose of this behaviour remains unclear.
It could be that the Dorkbot operators were changing servers to beat blacklisting services, or for some financial benefit.
However, whatever the cause, it is difficult to attribute this patterned behaviour to the takedown operation.
As the data shows, Dorkbot had already begun to rotate between servers just before the takedown occurred.
Even if the takedown was a factor, this rotating behaviour could also have been part of Dorkbot's distributed delivery architecture~\cite{theregister_dorkbot_nodate}, and perhaps the reason for its apparent resilience to the takedown attempt.
It should be noted that this (slow) rotation between servers is not the same as Fast Flux, the latter of which involves a single domain rotating between multiple IP addresses in a short period of time (\eg within minutes).

Notwithstanding, we also observed Dorkbot domains that flux between several IPs per day, such as \texttt{masterhossting7772.in} and \texttt{superstar7747.pw}.
Given that online sources have identified these domains as malicious,\footnote{\url{https://www.malwareurl.com/ns_listing.php?as=AS45945}} it is likely that these servers used Fast Flux.

Beyond its heavy use of particular IP addresses, the Dorkbot operation also utilises some domains from a mix of regions, as shown in Figures~\ref{fig:disrupting:operations_network_dyn}(a) and \ref{fig:disrupting:operations_network_dyn}(d).
This spread of servers is consistent with other research that identified the Dorkbot C\&C infrastructure to be distributed among a number of intercontinental regions~\cite{theregister_dorkbot_nodate}.
Given that the Dorkbot operation only used a few, particular servers to spearhead its delivery activities, it is probable that these other servers were held in reserve as back-up infrastructure.

\subsubsection{Upatre}
\label{sec:disrupting:operation_analysis:network:upatre}
The Upatre operation also exhibits an interesting progression of network activity, which, like the Dridex operation, can also be divided into a number of distinct stages, depending upon which network characteristic one is focusing.

In general, the Upatre operation experiences a rapid increase in network activity in the first few weeks (1st October--12th November) up until the arrest and seizure is carried out against the Dyre operation.
Specifically, looking at Upatre's use of download URLs in Figures~\ref{fig:disrupting:operations_network_dyn}(a)--(b), we see that during this period the Upatre malware tends to operate through download URLs with \texttt{.com} (and to a lesser extent, \texttt{.ms}) suffixes.
The most common effective second-level domains that it uses in this period are \texttt{ymail.com} (Yahoo! Mail) and \texttt{afx.ms}, which is a domain registered by Microsoft Corporation and known to be associated with Outlook Mail.\footnote{\url{https://whois.domaintools.com/afx.ms}}
This is consistent with the observation that Upatre is often delivered to victims through malicious email attachments~\cite{zemanacom_what_nodate,symantec_dyre_nodate,intelligence_dyre_nodate}.
During the same period, we observe Upatre's varied and progressive use of IPs from different countries, led by its use of servers in the United States, Germany (DE), France, and Ukraine (UA), as shown in Figure~\ref{fig:disrupting:operations_network_dyn}(d).
It is also interesting to note that, as we see in Figure~\ref{fig:disrupting:operations_network_dyn}(e), the Upatre operators ensure that their delivery servers are distributed among a number of countries.
Clearly, the Upatre operation was distributed through servers across multiple geographic regions, such as edge CDN servers for email services.
A simple query of the data confirms this as we find Upatre malware being linked to hundreds of region-specific subdomains of various email servers during this early period, such as \texttt{\{region\}\{integer\}.afx.ms} or \texttt{email\{integer\}.secureserver.net}.

After the takedown operation, Upatre's network activity rapidly decreases over a number of weeks (12th November--24th December).
As security researchers have noted~\cite{zemanacom_what_nodate,symantec_dyre_nodate, intelligence_dyre_nodate}, the Dyre malware heavily relied on the Upatre dropper malware as its main infection vector.
As such, the taking down of the Dyre operation could have plausibly led to some reverberations in the Upatre operation, perhaps due to some infrastructure being shared between the two.
However, as we will later see, this drop in Upatre network activity corresponds to a drop in Upatre binaries being downloaded onto victim computers.
Therefore, it remains unclear what causal links could exist between the takedown of the Dyre operation (an Upatre payload) and the subsequent drop in Upatre downloads (a dropper predominantly delivered through malicious email attachments).
Likewise, the question also remains: what infrastructure could have been shared between the two operations? 

As time goes on, we observe contrasting behaviours between Upatre's use of download URLs/domains and its use of IPs.
Namely, Upatre's use of IPs has a downward trend over the ensuing months (24th December--29th September), as shown in Figures~\ref{fig:disrupting:operations_network_dyn}(d)--(e).
On the other hand, as Figures~\ref{fig:disrupting:operations_network_dyn}(a)--(c) show, its use of download URLs and domains is quite stable for the first few months (24th December--14th April), but then suddenly increases and fluctuates at a raised level (14th April--29th September).
This behavioural disparity between Upatre's use of IPs and its use of download URLs/domains is interesting.
In particular, we observe a transition from the two metrics being quite strongly correlated at one stage (\ie their correlated peak and trough between 1st October--24th December) to them becoming increasingly incongruent as time goes on.\footnote{Pearson's and Spearman's correlation coefficients were computed for the Upatre IP count vs. FQDN count during three periods (inclusive): 1st October--24th December, 31st December--14th April, 21st April--29th September. $(r,\rho)$ as follows: Oct\_Dec$(0.76,0.60)$, Dec\_Apr$(0.63,0.45)$, Apr\_Sep$(0.41,0.11)$.}
This could indicate a significant change in Upatre's upstream delivery infrastructure some point after the Dyre takedown operation, such as a move from a distributed architecture to a more centralised one.
We find some evidence to support this hypothesis.
First, in Figures~\ref{fig:disrupting:operations_network_dyn}(a)--(b), we observe clear displacement in the Upatre operation from one set of domains to another:
particularly from sites with \texttt{.com} and \texttt{.ms} TLDs (such as \texttt{*.ymail.com} and \texttt{*.afx.ms}) to those with \texttt{.net} and \texttt{.ru} suffixes (such as \texttt{*.alfafile.net}).
Second, as we see in Figure~\ref{fig:disrupting:operations_network_dyn_2}(b), from around 14th April we observe an increase in the use of IPs that are associated with $2+$ e2LDs, corresponding to Upatre's migration to the \texttt{.net} and \texttt{.ru} domains.
Upon further inspection, these new servers (particularly those with \texttt{.ru} suffixes) were most likely generated by a DGA.
For instance, on 28th April, we identified 139 domains with a common domain structure: a static keyword for the subdomain, a random sequence of words and numbers for the second-level, and the \texttt{.ru} TLD (\eg \texttt{slingto.scene-root85.ru}, \texttt{slingto.robbusymyself.ru}, and \texttt{slingto.hanghandle.ru}).
Furthermore, these domains were all clustered around the same set of IPs, some of which involved over 10 different e2LDs per cluster.
We also note Upatre's heavy use of the \texttt{alfafile.net} (a file-hosting platform) and its various subdomains around this time, apparently replacing the email services and CDNs that it relied on several months before.
This marked change in delivery infrastructure by the Upatre operators shows a complete change in their \textit{modus operandi} (\ie from using compromised email services to using malicious domains with DGA as infection vectors), and could very well be evidence of an adaptation to previous takedowns.


\begin{figure}[t]
    \centering
    \includegraphics[width=1.05\linewidth]{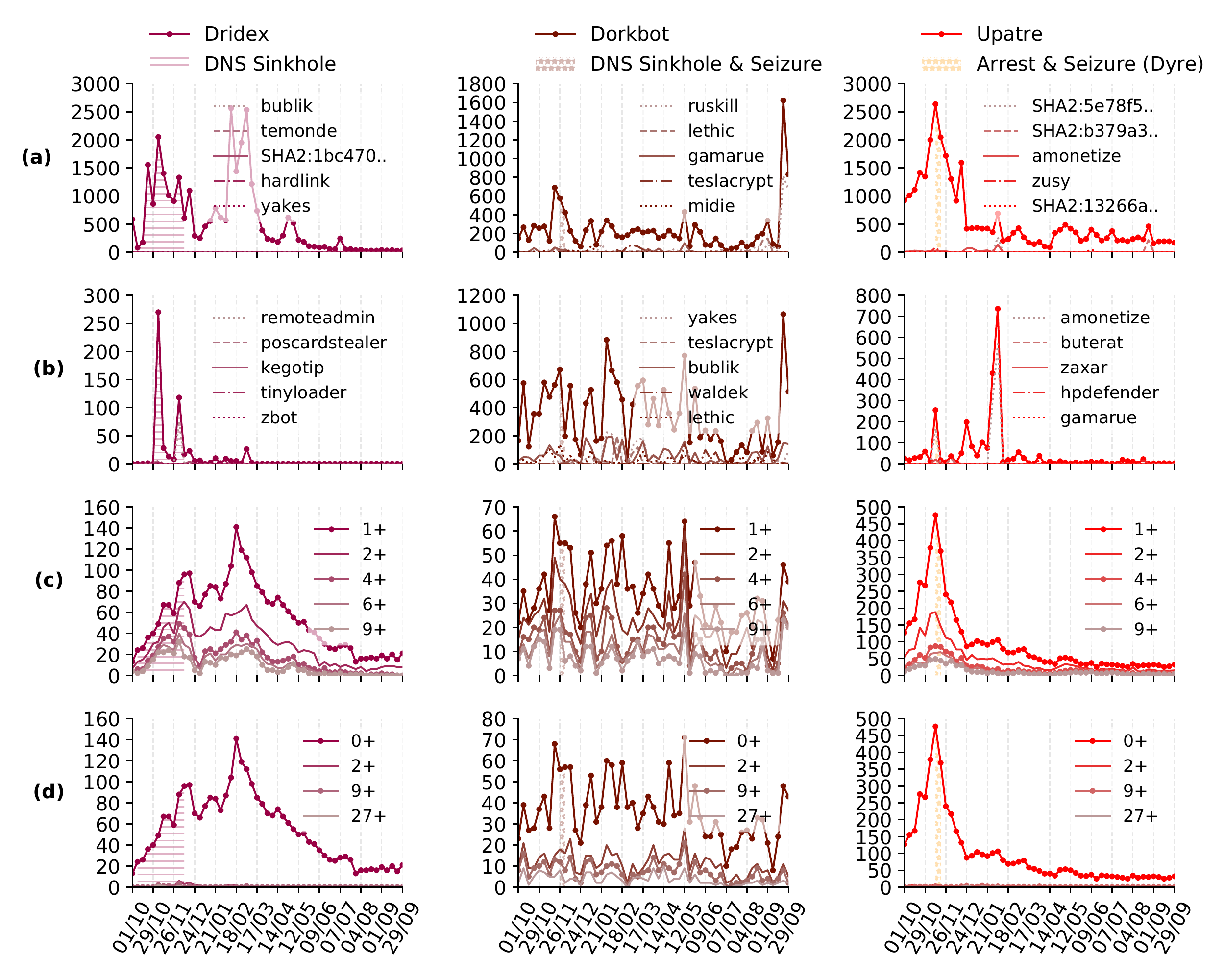}
    \caption{\textbf{Aggregate download activity:} \textbf{(a)} \# of times downloaded; \textbf{(b)} \# of drops by target malware; \textbf{(c)} \# of SHA-2s downloaded $N+$ times; \textbf{(d)} \# of SHA-2s that drop $N+$ files. Bursts of dropping activity by Dridex (during takedown) and Upatre (after takedown). Dorkbot activity more consistent throughout the year except for the sudden increase at the end. N.B: a few binaries are responsible for the majority of download activity (an approximate Power law relationship).}
    \label{fig:disrupting:operations_downloader_dyn}
\end{figure}

\subsection{Downloader Dynamics}
\label{sec:disrupting:downloader_dynamics}
In the last section, we analysed the network-level dynamics pertaining to each of the three malware delivery operations under study.
In this section, we move our analysis on to the characteristics and download activities of the malicious binaries themselves, which are fundamental to malware delivery operations.
In particular, we juxtapose the aggregate downloader dynamics, familial relationships (parent, children), delivery tactics, and polymorphic behaviours of the three malware operations.
Figure~\ref{fig:disrupting:operations_downloader_dyn} shows the aggregate download dynamics of each malware operation, while
Figure~\ref{fig:disrupting:operations_downloader_rel} shows their relational dynamics (\ie \# of parent and child files),
Figure~\ref{fig:disrupting:operations_downloader_eva} shows indicators of distributed delivery tactics, and Figure~\ref{fig:disrupting:operations_downloader_poly} indicators of polymorphic behaviour by the binaries.

\subsubsection{Aggregate download activity}
Figures~\ref{fig:disrupting:operations_downloader_dyn}(a)--(b) show the aggregate downloads and dropping activities of the Dridex, Dorkbot, and Upatre malware, whereas Figures~\ref{fig:disrupting:operations_downloader_dyn}(c)--(d) show the distributions of files that are downloaded $N+$ times, or drop $N+$ files.
One notices similar download behaviours between the Dridex and Upatre malware, but significantly different behaviours from Dorkbot.
This becomes a recurring theme in our analysis of download activities.

For the \textit{\textbf{Dridex}} malware, we observe ``bursts'' of downloads and dropping activity during the takedown counter-operation, and resurgence of (just) download activity between 
11th February--11th March, in correspondence with the peak in its network behaviours around the same time.
This supports the notion that the Dridex operators expanded their operation during the law enforcement takedown, perhaps in anticipation of (or in retaliation to) the expected disruptions due to the DNS sinkhole.
It is worth noting that that 95.8\% of the files dropped by Dridex between 29th October--24th December were unclassified.
Nonetheless, we identify a few instances of known malware families being delivered by Dridex, including some backdoor malware (\texttt{farfli}, \texttt{tinyloader}), financial fraud trojans (\texttt{zbot}, \texttt{zusy}, \texttt{poscardstealer}), among others (\texttt{troldesh}, \texttt{yakes}, \texttt{kegotip}).
It is difficult to draw any formidable conclusions on this aberrant behaviour given the lack of ground truth on the files dropped by the Dridex malware.
Still, it is interesting to see Dridex - a financial fraud trojan known to operate only as a payload rather than a dropper - suddenly engage in this practice of diversified, downstream malware delivery.
Looking at Figure~\ref{fig:disrupting:operations_downloader_dyn}(c), it appears (at least, visually) that the Pareto principle applies to the frequency of downloads for each Dridex file, where the majority are only downloaded once while decreasing proportions of files are downloaded more frequently.
On the other hand, as we see in Figure~\ref{fig:disrupting:operations_downloader_dyn}(d), almost none of the Dridex binaries engage in dropping activities.
Rather, through querying the data, we find that only up to 3 binaries are responsible for all dropping activity on any given day.
This supports the notion that the Dridex malware was primarily designed to operate as a malicious payload rather than an intermediate dropper.
However, specific strains of this malware were clearly modified to drop other malware onto victim systems.

With the \textit{\textbf{Upatre}} malware, we observe similarities to that of the Dridex malware.
As we see in Figure~\ref{fig:disrupting:operations_downloader_dyn}(a), and much like its network activity, we observe a peak in Upatre downloads just before the arrest and seizure counter-operation around 19th November.
We also observe several ``bursts'' of Upatre dropping activity in Figure~\ref{fig:disrupting:operations_downloader_dyn}.
Of the files that Upatre drops, we find that on 12th November, 60\% were PUP (mostly \texttt{convertad}) and 23\% malware; on 24th December, 98\% were unclassified; and between 28th January--4th February, 77\% were PUP (mostly \texttt{amonetize}) and 3\% malware.
It is interesting to see that such a high proportion of Upatre payloads are PUP such as \texttt{convertad} and \texttt{amonetize} (as opposed to other malware), which are families known to bundle and integrate with legitimate software.\footnote{\url{https://www.shouldiremoveit.com/ConvertAd-88792-program.aspx}}
This case study gives an indication of how convoluted file dependencies and delivery chains between malware, PUP, and benignware can be in the wild. 
As we look at the bounded frequency plots of downloads per SHA-2 and drops per SHA-2 in Figures~\ref{fig:disrupting:operations_downloader_dyn}(c)--(d), we see a similar case as with the Dridex malware: (i) an apparent, inverse relationship between SHA-2 count and the frequency in which each SHA-2 is downloaded; and (ii) a minority of files being responsible for all of the Upatre's dropping activity.
The latter observation is more strange in this case, given that Upatre is known to operate mainly as a dropper malware.
More generally, we find that Upatre is downloaded more frequently than it downloads other files within this observation window.

Analysing the \textit{\textbf{Dorkbot}} malware, we observe significantly different download behaviours than the other malware families.
First, as we see in Figures~\ref{fig:disrupting:operations_downloader_dyn}(a)--(b), the download and dropping dynamics of the Dorkbot operation do not appear to change significantly over the course of the year (including the takedown period), barring a sudden increase at the end of the observation period.
We previously noted that it was difficult to attribute Dorkbot's ever-changing network behaviours to the takedown counter-operation.
The lack of any significant change in Dorkbot's overall download activity over the observation period seems to support this position even further.
In Figure~\ref{fig:disrupting:operations_downloader_dyn}(c), the plots of downloads per SHA-2 for the Dorkbot malware show a generally ``flatter'' distribution between each group (\ie more evenly spaced plots for $N=1,2,3,...$).
This seems to indicate a weaker Pareto distribution (if any) in comparison to the other operations.
The Dorkbot operation is also differentiated by its higher proportion of file SHA-2s that engage in dropping behaviour.
Specifically, in Figure~\ref{fig:disrupting:operations_downloader_dyn}(d), while most do not engage in any dropping behaviours, up to 40\% of Dorkbot SHA-2s deliver $9+$ subsequent payloads over the course of the observation period.

\subsubsection{Relational dynamics}
In Figure~\ref{fig:disrupting:operations_downloader_dyn} we observed the aggregate download activity of the three malware operations under study.
It is also important to understand the other software families that contribute to this activity, either as droppers (\ie parent files that download the target malware), or as payloads (\ie child files that are dropped by the target malware).
In particular, Figures~\ref{fig:disrupting:operations_downloader_dyn}(a)--(b) show the top 5 labelled software that either download the target malware (parent files) or are downloaded by the target malware (child files).
In most cases, we see that these ``top'' families account for a very small percentage of the overall download activity of the target families.
The exception to this appears to be the case of the \textbf{\textit{Dorkbot}} operation, where in Figure~\ref{fig:disrupting:operations_downloader_dyn}(a) we see a sharp increase in \texttt{ruskill} downloads towards the very end of the observation window, while in Figure~\ref{fig:disrupting:operations_downloader_dyn}(b) we see that the \texttt{yakes}, \texttt{teslacrypt}, and \texttt{bublik} malware families account for most of Dorkbot's dropping activities.

\begin{figure}[!htbp]
    \centering
    \includegraphics[width=1.05\linewidth]{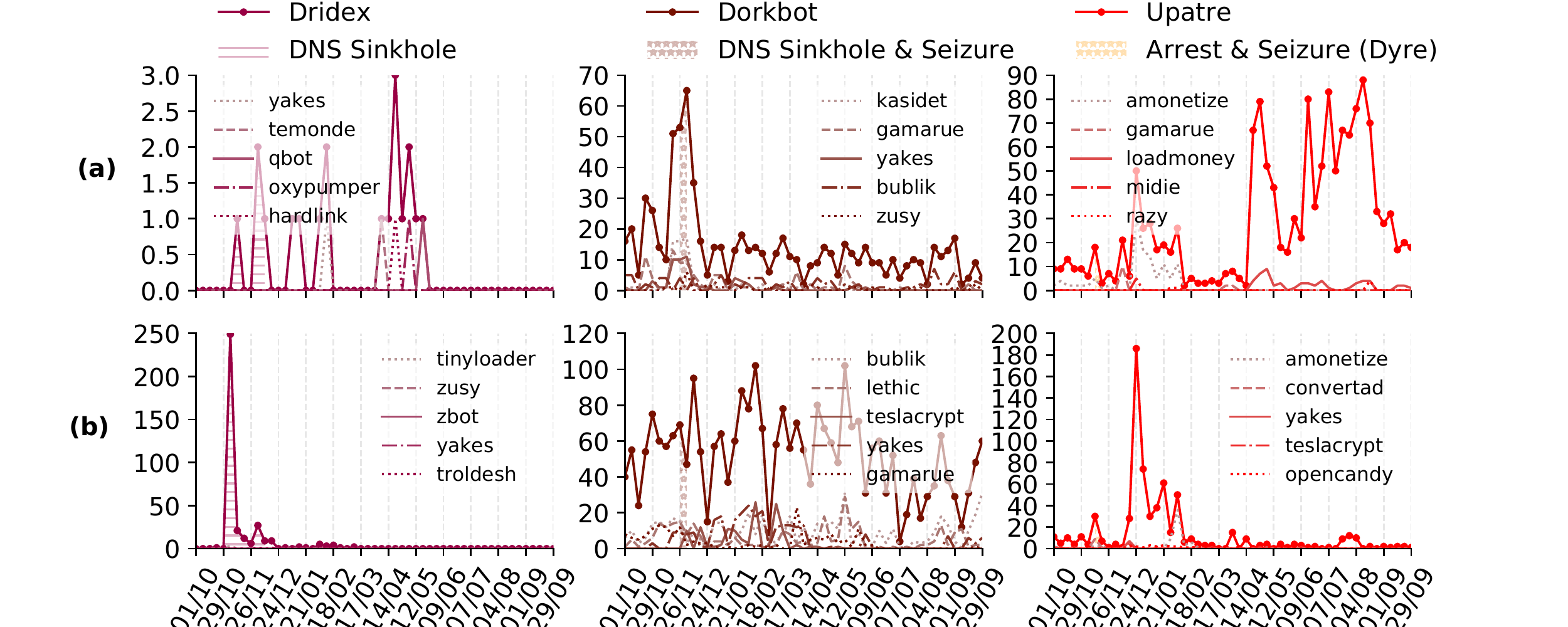}
    \caption{\textbf{Relational dynamics:} \textbf{(a)} \# of SHA-2s that download target malware; and \textbf{(b)} \# of SHA-2s dropped by target. N.B: the sharp increase in Upatre upstream droppers after mid-April, correlating with its increased use of DGA servers.}
    \label{fig:disrupting:operations_downloader_rel}
    \includegraphics[width=1.05\linewidth]{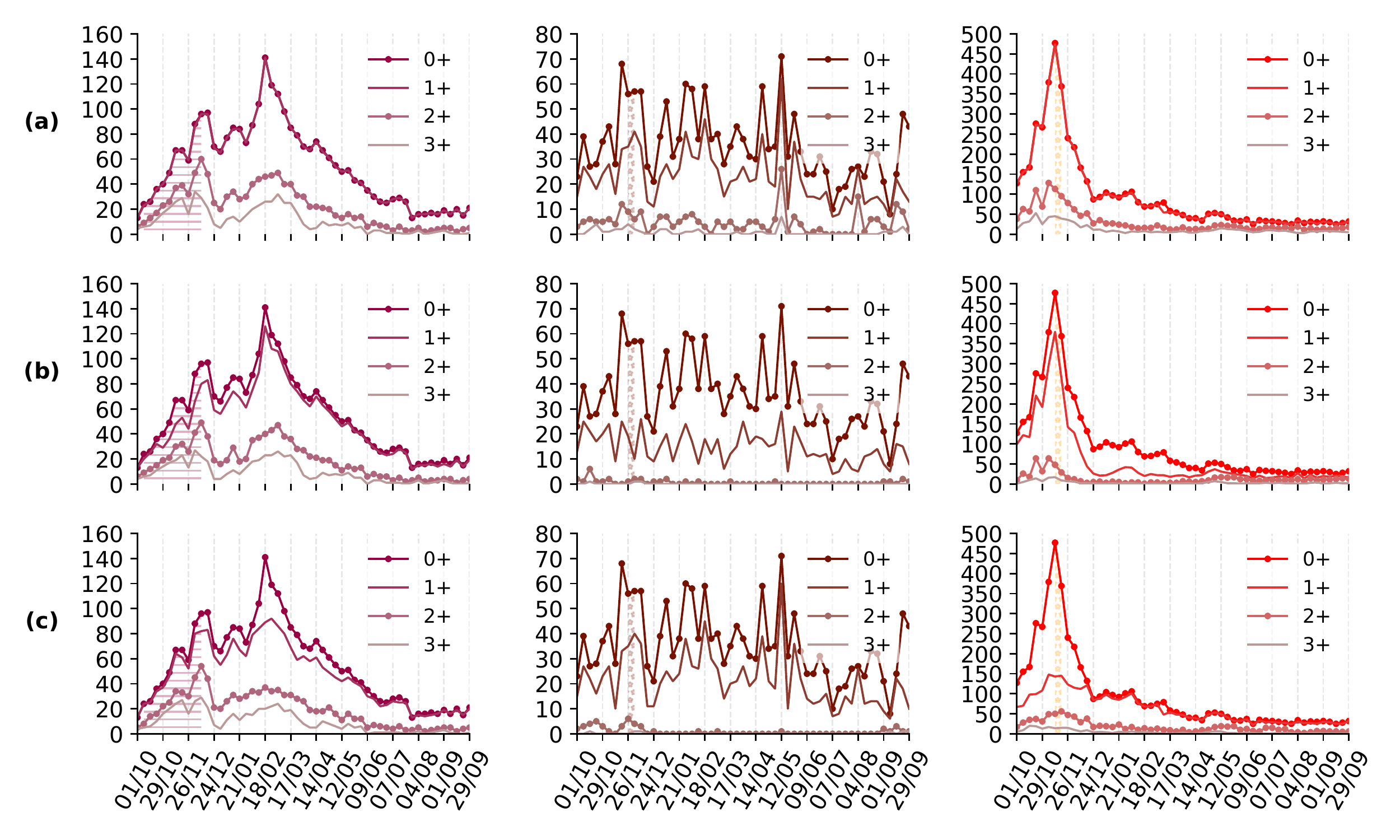}
    \caption{\textbf{Distributed delivery indicators:} \textbf{(a)} \# of SHA-2s associated with $N+$ URLs; \textbf{(b)} \# of SHA-2s associated with $N+$ e2LDs; and \textbf{(c)} \# of SHA-2s associated with $N+$ IPs. Dorkbot downloads often without any traceable network resource, alluding to direct writing to filesystems. }
    \label{fig:disrupting:operations_downloader_eva}
    \includegraphics[width=1.05\linewidth]{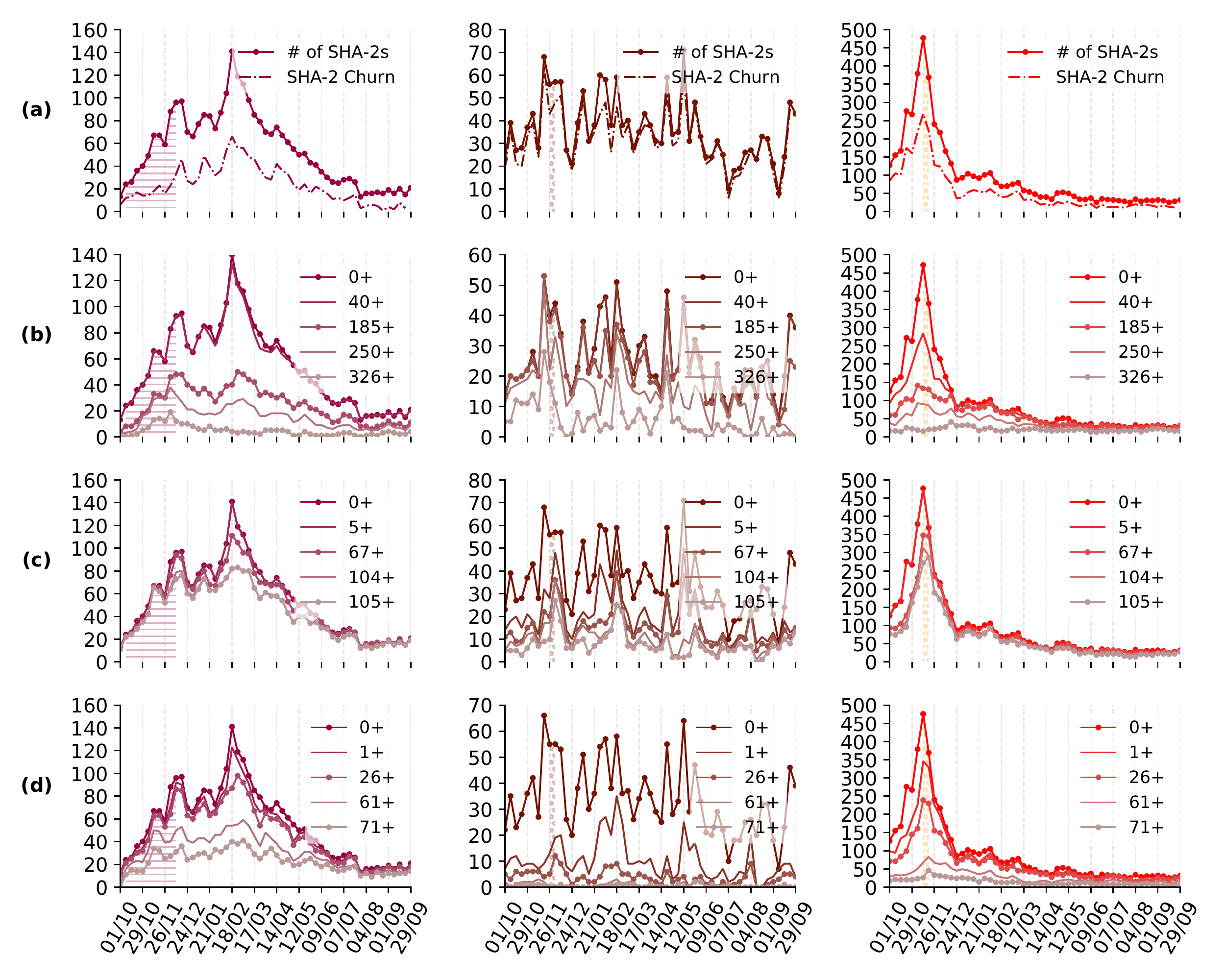}
    \caption{\textbf{Polymorphic characteristics:} \textbf{(a)} \# of active SHA-2s and SHA-2 churn; \textbf{(b)} \# of SHA-2s of size $N+$ KB; \textbf{(c)} \# of SHA-2s with $M+$ malice score, where $0 \geq M \geq 128$; and \textbf{(d)} \# of SHA-2s with $P+$ prevalence score, where $0 \geq P \geq 127$.\\
    N.B: malice and prevalence scores are assigned by Symantec.}
    \label{fig:disrupting:operations_downloader_poly}
\end{figure}

Turning to the question of how many families are related to the studied malware, Figure~\ref{fig:disrupting:operations_downloader_rel} shows the aggregate number of families involved in each malware operation.
For the \textbf{\textit{Dridex}} operation, Figure~\ref{fig:disrupting:operations_downloader_rel}(a) shows very few upstream malware distributing it during the year.
This implies that the Dridex operation relied more on server delivery infrastructure than dropper malware, which is consistent with other observations of this malware being delivered through malicious email attachments and exploit kit downloads~\cite{malwarebytes_trojandridex_nodate}.

The \textbf{\textit{Dorkbot}} behaves very differently.
As Figure~\ref{fig:disrupting:operations_downloader_rel}(a) shows, the Dorkbot malware relies consistently (of a cyclic nature) on upstream malware droppers.
Particularly up until the takedown, Dorkbot was delivered by malware such as \texttt{gamarue}, \texttt{kasidet}, and \texttt{yakes}.
However, after the takedown, the number of upstream malware in the Dorkbot operation dropped significantly, though, as previously noted, it's overall download activities seemed unaffected for the most part.
Given the lack of ground-truth in this regard, it is difficult to ascertain whether the takedown only affected a subset of the Dorkbot operation (\ie upstream dropper networks).
In like manner, we see that Dorkbot also distributed a wide range of downstream malware throughout the observation period.
Again, one cannot see any sign of diminished activity due to the takedown.

The \textbf{\textit{Upatre}} operation also exhibits some interesting relational behaviours.
In particular, as Figure~\ref{fig:disrupting:operations_downloader_rel}(a) shows, Upatre relies mostly on a few families in the first half of the observation window, such as the \texttt{amonetize} PUP and \texttt{gamarue} malware.
However, in the second half of the observation window, we see a significant change in behaviour: Upatre shifts to a diversified, upstream dropper network, as indicated by (i) a large increase in the total upstream families, and (ii) the ``top'' families (\eg \texttt{loadmoney}) accounting for only a small proportion of them.
Though it is unclear why the cause of this change, we note it occurring from 14th April onwards -- the same time Upatre began to use DGA download servers (see Section~\ref{sec:disrupting:operation_analysis:network:upatre}).

\subsubsection{Distributed delivery tactics}
Figure~\ref{fig:disrupting:operations_downloader_eva} shows distributed delivery metrics of each malware operation: the numbers of SHA-2s
associated with varying numbers of URLs, e2LDs, and IPs.
Again, we observe similarities in the Dridex and Upatre operations, but considerably different characteristics in the Dorkbot operation.

Starting with the bounded frequency plots of \textbf{\textit{URLs per SHA-2}} in Figure~\ref{fig:disrupting:operations_downloader_eva}(a), we see that almost all the download activities of the Dridex and Upatre SHA-2s involve network activity, as indicated by the near-total overlap of the plot lines for $N=0$ and $N=1$.
This is in stark contrast to the Dorkbot malware, which shows significant ``gaps'' between the $N=0$ and $N=1$ plot lines, indicating that some files are not associated with any download URL.
This could allude to Dorkbot writing directly to the filesystem from the malicious process as opposed to initiating the download from an external server.
This is consistent with reports which identified spreading through USB flash drives as one of Dorkbot's infection vectors~\cite{microsoft_win32dorkbot_nodate}.
It is still possible, however unlikely, that this discrepancy could be due to some measurement error in the data collection process.
Nonetheless, we see that SHA-2s being associated with multiple URLs is a common occurrence for these malware operations (although relatively less common for the Dorkbot operation).

Figure~\ref{fig:disrupting:operations_downloader_eva}(b) shows the bounded frequency plots of \textbf{\textit{e2LDs per SHA-2}}, while
Figure~\ref{fig:disrupting:operations_downloader_eva}(c) shows \textbf{\textit{IPs per SHA-2.}}
Most of the Dridex malware is associated with at least one e2LD or an IP, while up to 50-60\% of its files are associated with $2+$ e2LDs/IPs.
It is particularly interesting to see that the highest proportion of files associated with multiple e2LDs/IPs occurred during the takedown period.
Again, this supports the notion that a concerted effort was made by Dridex operators to ramp up malware activity during the sinkhole operation.
The Upatre operation exhibits significantly different characteristics: the proportion of its files that are associated with $1+$ e2LDs/IPs is highly variable across the year.
For instance, between 1st October--24th December, there is a significant evolution in its delivery patterns: (i) a sudden rise and fall in files associated with $1+$ e2LDs, and (ii) at one point, the majority of files having no traceable IP.
It remains unclear why Symantec's telemetry could not detect IPs for these download events, or why these files were prominent only in the early part of the observation window.
Nonetheless, it is unlikely this was a random occurrence, given these correlated behaviours.
Finally, the Dorkbot malware exhibits much of the same delivery patterns as before: a significant (but still minor) proportion of its files are not linked to any server.
This alludes to binaries writing directly onto victim filesystems.

\begin{table*}[t]
\centering
\resizebox{\linewidth}{!}{
\begin{tabular}{p{1.5cm}|p{5cm}|p{5cm}|p{5cm}}
\toprule
 & \textbf{\textit{Dridex}} & \textbf{\textit{Dorkbot}} & \textbf{\textit{Upatre}} \\
\hline
\hline

\footnotesize \textit{\textbf{LEA takedown}} 
& \footnotesize 60-day DNS Sinkhole and Disinfection. 
& \footnotesize DNS Sinkhole and Seizure. 
& \footnotesize Arrest and Seizure. \\

\hline

\footnotesize \textit{\textbf{Malware operation behaviours}} 
&
\footnotesize $\bullet$ Malware operations increase and diversify during first half of observation window (including LEA takedown). Gradually decreases in second half of window. 
\newline $\bullet$ Distributed delivery architecture: significant use of shared-hosting platforms and multi-region CDNs.
\newline $\bullet$ Sparse bursts of dropping activity: delivered other malware including ransomware, banking trojans, backdoors. Uncharacteristic of Dridex malware.
\newline $\bullet$ Minority of files responsible for majority of downloads / all dropping activity.
\newline $\bullet$ Few upstream droppers; heavy reliance on upstream network infrastructure.
\newline $\bullet$ Up to 60\% files delivered by $2+$ e2LDs/IPs.
\newline $\bullet$ Significant polymorphism and churn rate (up to 60\%). High detection rates (prevalence/malice scores).

& 
\footnotesize $\bullet$ Highly cyclic/stochastic operational activity.
\newline $\bullet$ Distributed delivery architecture: multi-region servers. 
\newline $\bullet$ Coordinated rotation between servers in different countries over observation window. Likely use of Fast Flux also.
\newline $\bullet$ Sharp but brief drop in network activity after LEA takedown. No observable long-term effects.
\newline $\bullet$ Potentially held back-up infrastructure.
\newline $\bullet$ Slightly ``flatter'' distribution of download activity across SHA-2s.
\newline $\bullet$ Sharp increase in downloads at end of observation window: mainly delivered by \texttt{ruskill}.
\newline $\bullet$ Consistent reliance on upstream droppers; mixed reliance on upstream network infrastructure.
\newline $\bullet$ Broad range of downstream malware dropped.
\newline $\bullet$ Likely use of direct writing to file system (\eg binary replication).
\newline $\bullet$ Extremely high polymorphism and churn rate (almost 100\%). Low-to-mild detection rates (prevalence/malice scores).

&
\footnotesize $\bullet$ Rapid, initial increase in operational activity; sharp drop after takedown.
\newline $\bullet$ High use of email services (initially) and IPs in multiple regions.
\newline $\bullet$ Apparent shift in delivery infrastructure over observation window: distributed to more centralised.
\newline $\bullet$ Displacement in domains used (from \texttt{.com} and \texttt{.ms} to \texttt{.ru} and \texttt{.net}).
\newline $\bullet$ Increased use of DGA servers in latter half of window; corresponding decreased use of mail servers.
\newline $\bullet$ Dropped a range of downstream software in bursts: mainly PUP; some malware and unlabelled families.
\newline $\bullet$ Minority of files responsible for majority of downloads / all dropping activity.
\newline $\bullet$ Relies on a few upstream droppers in first half of window; sudden change and increase of upstream droppers in second half (correlated with DGA usage).
\newline $\bullet$ Significant reliance on upstream network infrastructure.
\newline $\bullet$ Significant polymorphism and churn rate (up to 80\%). Mild-to-high detection rates (prevalence/malice scores). \\

\bottomrule
\end{tabular}
}
\caption{Summary of LEA takedowns and observed behaviours of the targeted malware delivery operations.}
\label{tab:summary_malware_behaviours}
\end{table*}

\subsubsection{Polymorphism}
Figure~\ref{fig:disrupting:operations_downloader_poly} shows the polymorphic characteristics of each malware operation.
The \textbf{\textit{number of active SHA-2s}} (or malware variants) and \textbf{\textit{churn rates}} for each malware are shown in Figure~\ref{fig:disrupting:operations_downloader_poly}(a).
Clearly, each of the malware delivery operations makes extensive use of polymorphism during the observation window.
Furthermore, we see that the active SHA-2 count of each malware evolves much like the network dynamics of its respective delivery operation.
For example, the active SHA-2 count for the Dridex operation increases while the DNS sinkhole takes place, and falls some months after; that of Upatre falls sharply after the arrest and seizure occurs (although its network components behave very differently in the second half of the observation window); that of the Dorkbot operation continues to fluctuate in apparent immunity to its respective takedown.
This correlation in SHA-2 count and the number of network components used to deliver them (URLs, domains, IPs)\footnote{Pearson's and Spearman's correlation coefficients were computed for SHA-2 count vs. URL count over 1st October--14th April -- the period for which these relationships are approximately linear. $(r,\rho)$ as follows: Dridex$(0.93,0.93)$, Dorkbot$(0.74,0.71)$, Upatre$(0.99,0.93)$.}
could be the result of campaign IDs being hard-coded into each binary, being unique to each upstream distributor.
In this case, the binaries delivered by each distributor would have a different file hash.
Looking at the churn rates, we see that all of the operations exhibit high churn.
Nonetheless, Dorkbot exhibits exceptionally higher churn rates, where almost all its SHA-2s are replaced weekly.

Figure~\ref{fig:disrupting:operations_downloader_poly}(b) shows the distribution of \textbf{\textit{file sizes}} (in KB).
We observe significant variability in the sizes of each malware, although most SHA-2s are less than 326KB.
It should be noted, however, a few binaries as large as 15MB were observed in the data (particularly Upatre binaries).
It is unclear whether this variability in file size (or how much of it) is a result of some polymorphic technique (\eg binary padding), or if it's simply due to additional functionality being coded into certain versions of these malware.

Figure~\ref{fig:disrupting:operations_downloader_poly}(c) shows the distribution of assigned \textbf{\textit{malice scores}}, while Figure~\ref{fig:disrupting:operations_downloader_poly}(d) shows the distribution of \textbf{\textit{prevalence scores}}.
It is interesting to see that most Dridex and Upatre SHA-2s are assigned very high malice scores with very low variance, while Dorkbot is assigned much more variable malice scores.
This suggests that Dorkbot was much more successful than the other malware at evading detection systems such as Symantec and the other antivirus engines used to generate these scores.
Likewise, Dorkbot is generally assigned much lower prevalence scores than the other malware.
This indicates that the detection systems did not observe Dorkbot malware as frequently at the time.
This is most likely the result of Dorkbot's very high churn rate, which could also be a contributing factor to it being assigned significantly lower malice scores.

\subsection*{Summary of Results}
\label{sec:disrupting:summary_results}

In this section, we presented a comprehensive analysis of the network and download activities of three, different malware delivery operations, and how they evolved over the course of a year in light of law enforcement efforts to disrupt them.
A summary of these observations is presented in Table~\ref{tab:summary_malware_behaviours}.

%% file: discussion.tex
\section{Discussion}
\label{sec:disrupting:discussion}

We conducted a detailed analysis of the dynamics and behaviours of three malware delivery operations over the course of a year.
In this section, we take a step back to consider the implications of these findings.
Specifically, we identify what the security community can learn from these observations, and how these findings could be factored into future countermeasures.
We also reflect on the limitations of this study, and opportunities for future work.

\subsection{Lessons Learned}
\label{sec:disrupting:discussion:lessons}
We observed a diversity of structural designs, behaviours, patterns, and responses to takedown attempts in the studied operations, finding the following commonalities between them.

\subsubsection{Distributed delivery architectures}
All three operations made significant use of distributed delivery infrastructures: Dridex used shared-hosting services and CDNs in up to 35 different countries; Dorkbot constantly rotated between international servers; and Upatre heavily used multi-region CDNs and cloud services (\texttt{ymail.com}, \texttt{alfafile.net}).
This has been observed of malicious file delivery operations in multiple studies~\cite{ife2019waves,lever2017lustrum,dittrich2012so,stringhini2017marmite}.
This makes effective server-based takedowns more difficult, thus requiring greater coordination between LEAs, security companies, and service providers on the Internet.
Most especially, given that these service providers have been so commonly abused, it is pertinent that they continue to step up their security hygiene and coordination with other stakeholders to prevent cybercriminals from abusing such platforms.

\subsubsection{Polymorphism and Pareto's principle}
Polymorphism was rigorously employed by all three malware.
However, some malware binaries (Dridex, Upatre) were detected more frequently than others (Dorkbot).
One possible explanation for this is that a malware (such as Dorkbot) that churns through binaries more frequently would be more difficult to detect in the short-term.
We also observed manifestations of Pareto's principle across all malware operations in that a minority of binaries were responsible for a majority of download activity.
Although detecting polymorphic malware will be a continued challenge for the security community, this skewed distribution of activity towards a minority of binaries indicates that detecting these ``super binaries'' would yield the most benefit.

\subsubsection{Takedown resilience}
Each malware operation responded differently and showed some degree of resilience to takedown.
For instance, Upatre shifted to a more centralised infrastructure over several months; Dridex significantly increased its activity \textit{during} the LEA takedown attempt; Dorkbot showed no significant changes, but continued in it's cyclic/stochastic behaviours and likely use of Fast Flux.
In view of this, one may ask the age-old question of whether botnet takedowns are \textit{actually} effective?
Researchers have found that, historically, the success of botnet takedowns is highly variable~\cite{stone2011underground,edwards2015analyzing}.
Perhaps a more pertinent question to ask is whether botnet takedowns are the \textit{only} effective means to controlling malware delivery?
Granted, there are alternative takedown techniques that could also be employed, such as infiltrating botnet infrastructure and disrupting them from within~\cite{stone2009your,binsalleeh2010analysis,eshete2015ekhunter}.
However, by viewing malware delivery as a supply chain problem, for example, the security community may achieve more success by targeting other aspects of the malware economy in parallel, such as by attacking the flow of money around malware delivery (the reliability of Dark markets, the process of monetising stolen data and compromised devices, \etc).
It has also been argued~\cite{ife2019bridging} that the security community could elicit more disruptive techniques from other fields of security research. 
For example, frameworks such as Situational Crime Prevention~\cite{clarke_situational_1997} could be adapted to derive countermeasures against botnet operations~\cite{ife2019bridging}.

\subsubsection{Predictable responses}
Environmental criminology literature recognises several types of offender responses to anti-crime interventions.
These include (i) \textit{displacement} -- a change in an offender's behaviour to circumvent the intervention or seek out alternative targets or crime types~\cite{hesseling_displacement_nodate}; (ii) \textit{adaptation} -- a longer-term process of displacement whereby the offender population as a whole discover new crime vulnerabilities and opportunities after an intervention has been in place for a while~\cite{ekblom_gearing_1997}; and (iii) \textit{defiance} -- an increase in offender activity in retaliation to an intervention, usually when the offender perceives the intervention as unjust or disproportionate~\cite{sherman_defiance_1993}.
Behaviours such as these are usually expected and factored into interventions supported by environmental criminology.

Similarly, in this study, we observed interesting responses from the malware operators to takedown efforts.
For instance, the Dridex operators significantly ramped up botnet activity during the DNS sinkhole counter-operation, with an increased concentration of servers in the US and UK.
We noted that this was the second or third LEA counter-operation against the Dridex botnet in as many months.
Assuming this is linked to the attempted takedowns, this is characteristic of defiant and displacing behaviours.
Likewise, we observed significant changes in the Upatre infrastructure only a few months after the Dyre takedown.
Particularly, it shifted in its use of multi-region email services to more centralised clusters of DGA servers and a single CDN (\texttt{alfafile.net}).
Again, this is characteristic of displacement, potentially to regain more control of the malware delivery process.

As such, the main takeaway here is that, much like crime in the physical world, reactions from the malware operators must be expected and factored into any mitigation strategy against their operations.
This highlights the importance of two things: first, the continued monitoring and management of malware operations, before, during, and after any takedown attempt (\eg assessing the potential for unwanted side-effects~\cite{chua2019identifying}, implementing action-research models for botnet takedowns~\cite{ife2019bridging}); and second, the necessity for security researchers, companies, and LEAs to disseminate information regarding botnet takedown attempts, as this shared body of knowledge would better equip the security community to implement effective countermeasures.
Nonetheless, there is the argument that cybercriminals could also learn how to make their operations more resilient through this shared knowledge.
This raises the question of how best to implement such knowledge-sharing.

\subsubsection{Unpredictable responses}
At the same time, we also observed very \textit{aberrant} and previously undocumented behaviours by each malware operation.
For instance, though Dridex is a financial fraud trojan and has been known to operate as a payload, we observed it engaging in bursts of dropping activity, delivering downstream ransomware, backdoor malware, and even competing families of financial fraud trojans!
Dorkbot exhibited sudden and sharp increases in downloads at the end of the observation period through upstream \texttt{ruskill} malware.
Upatre suddenly and significantly increased in its use of upstream malware droppers in the latter half of the observation period.
Such behaviours could be very difficult to predict, especially when monitoring malware activity from a limited perspective (\ie download traffic).
This highlights the need for the security community to incorporate data sources from multiple ecosystems to monitor botnet activity effectively.
For instance, monitoring download traffic (as in this study) could be complemented with other intelligence sources, such as network traffic from ISPs, online discussions in social media and web forums (Twitter, Reddit), as well as discussions and market activity in the Dark Web.
Potentially, using multiple perspectives could give researchers more context and clarity regarding some of these observed behaviours.

\subsection{Limitations}
\label{sec:disrupting:discussion:limitations}

This work builds on the data and techniques used in a previous measurement study of the malicious file delivery ecosystem~\cite{ife2019waves}.
As such, the same data limitations apply, such as the limited view we have on only one stage of the malware supply chain (software download), or VirusTotal's limited coverage in mappings between file hashes and malware families.
To mitigate the former issue, we used additional data sources to provide as much context as possible (ground truth on the operations, VirusTotal/AVClass/NSRL software labels, malware aliases, \etc). 
To mitigate the latter issue, we collected VirusTotal labels for a period of three years after the initial observations, maximising positive predictive performance.
It is still possible that some files were mislabelled with the wrong malware family, which would mean that the time series analytics is unrepresentative of the given family. However, we suspect such cases would be few given the reported accuracy of the classifier~\cite{sebastian2016avclass}.

A major part of this study involved analysing malware delivery operations that were subject (or in the case of Upatre, linked) to a takedown attempt.
However, a number of challenges arise.
One challenge relates to the fact that ground truth on takedown operations is usually scarce. This was the case with the operations studied herein.
As such, this study is limited regarding the specifics of each takedown operation, and finding parallels in the data.
More generally, and as a result of this general lack of ground truth data on takedown operations, this study was scoped as a measurement study of global malware activity.
This means that we are only able to observe and evaluate the overall structure and activities of each malware operation but cannot do more than speculate why such phenomena occur, nor can we isolate observable effects to the specific parts of each infrastructure that were targeted for takedown.
In light of this challenge, one interesting extension to this work could be the use of a causal inference framework to analyse the effects of takedown attempts on different aspects of each malware operation (aggregate network and download activity, distributed delivery, \etc), as well as the wider malicious file delivery ecosystem.
Alternatively, causal relationships could be uncovered more directly with additional ground truth on the specifics of each takedown operation.
Another, more general challenge is the issue of \textit{survivorship bias}.
In the context of this work, this refers to the biases that arise out of the fact that certain characteristics of the studied botnets would make them more likely to be targeted for takedown than other botnets. Such biases ultimately threaten the external validity of these findings (\ie how well they apply to other botnets, particularly those not targeted for takedowns).

Finally, on the topic of understanding the behaviours of the malware operators, it is also worth noting that we could only observe \textit{spatial displacement} in this study (\ie an operator moving from one set of upstream servers and dropper networks to another).
The methodology could be extended to include \textit{ecosystem dynamics} that could allow us to observe \textit{offender displacement} (\ie a malicious operator replacing another's use of upstream delivery infrastructure).

%% file: conclusion.tex
\section{Conclusion}
\label{sec:disrupting:conclusion}

In this study, we tracked and analysed three different malware delivery operations over the course of a year, studying the dynamics of their upstream servers and dropper networks.
We made a number of key findings -- mainly, the tendency of malware operators to move their operations elsewhere after a takedown, or in one case, to openly defy it.
We also found the use of distributed delivery architectures (particularly CDNs) and the heavy reliance on a few ``super binaries'' to be common by the studied malware operators.
These observations give the security community deeper insight into the complexities of malware delivery and ought to be factored into future takedown strategies.


%% file: acknowledgements.tex
We would like to thank the anonymous reviewers for their comments on this work.
Colin C. Ife was supported in part by EPSRC under grant no. EP/M507970/1.
Steven J. Murdoch is supported by The Royal Society under grant no. UF160505.
Gianluca Stringhini is supported by NSF under grant no. CNS-1942610.